\documentclass[%
reprint,
amsmath,amssymb,
aps,
floatfix,
]{revtex4-2}

\usepackage{graphicx}
\usepackage{dcolumn}
\usepackage{bm, parskip}

\usepackage{xcolor}

\usepackage[normalem]{ulem}

\begin{document}

\preprint{APS/123-QED}

\title{Emergent ordering in active fluids driven by substrate deformations:\\ Mechanisms and patterning regimes}

\author{Varun Venkatesh}

\author{Amin Doostmohammadi}%
 \email{doostmohammadi@nbi.ku.dk}
\affiliation{%
 Niels Bohr Institute, University of Copenhagen\\
 Blegdamsvej 17, 2100 Copenhagen, Denmark 
}%

\date{\today}

\begin{abstract}
The interplay between active matter and its environment is central to understanding emergent behavior in biological and synthetic systems. Here, we show that coupling active nematic flows to small-amplitude deformations of a compliant substrate can fundamentally reorganize the system’s dynamics. Using a model that combines active nematohydrodynamics with substrate mechanics, we find that contractile active nematics - normally disordered in flat geometries - undergo a sharp transition to an ordered state when the environment is deformable. This environmentally induced ordering is robust and enables distinct patterning regimes, with wrinkle morphologies reflecting the nature of the active stresses. Our results reveal a generic mechanism by which mechanical feedback from soft environments can lead to ordering in active systems. 

\end{abstract}

\maketitle


Orientational order is one of central organizing principles in both living and synthetic systems. From collective cell migration to cytoskeletal alignment and active colloidal flows, the emergence of polar, nematic, or higher-order symmetries enables large-scale coordination and function across scales ~\cite{trepat_mesoscale_2018, huebsch_collective_2022,andersen2025evidence}. In passive systems, such order arises through thermodynamic phase transitions, such as the isotropic-to-nematic transition driven by density or temperature ~\cite{chaikin_principles_1995}. In active systems, by contrast, order emerges from the interplay between internal stresses and external constraints, and can arise far from equilibrium.

Active materials often generate dipolar stresses, broadly classified as extensile or contractile depending on whether they push or pull along their axis of alignment~\cite{pedley1992hydrodynamic,guasto2010oscillatory,drescher2010direct}. These stress types underlie diverse behavior, from microtubule–motor mixtures to actomyosin networks \cite{sanchez_spontaneous_2012, thoresen2011reconstitution}. While extensile systems have been shown to spontaneously develop nematic order through activity-induced alignment and flow feedback ~\cite{thampi_intrinsic_2015, hsu_activity-induced_2022, pearce_activity_2019}, contractile systems - typical of many cellular environments - remain isotropic in the absence of thermodynamic ordering cues. That is, contractile activity alone has so far been shown to be insufficient to break rotational symmetry and induce orientational order in the absence of thermodynamic ordering mechanisms \cite{saintillan2008instabilities,thampi_intrinsic_2015}. 

Several mechanisms have been proposed to organize active nematics, including flow-alignment under shear ~\cite{muhuri_shear-flowinduced_2007, mandal_shear-induced_2021}, confinement ~\cite{you_confinement-induced_2021}, and coupling to structured or compliant environments ~\cite{bischofs_cell_2003, callens_substrate_2020,plan2021activity}. Yet these pathways overwhelmingly support ordering in extensile systems, leaving open the question of how robust oreintational order arises in the contractile regime - especially given that contractility dominates in many biological settings.

Experimental evidence suggests that externally imposed mechanical cues can orient individual cells and filaments. Substrate curvature has been shown to guide cellular alignment and flow patterns ~\cite{bell_active_2022}, while microscale deformations influence organization via contact guidance and stiffness sensing ~\cite{lacroix_emergence_2024, sunyer_durotaxis_2020}. Cells also deform their environment, generating small-amplitude wrinkles not only on substrates~\cite{guillamat_guidance_2025} but within tissues and intracellular structures, such as epithelia ~\cite{andrensek_wrinkling_2023, youn_tissue-scale_2024}, microtubules ~\cite{strubing_wrinkling_2020}, or nuclei ~\cite{jackson_scaling_2023, wang_extreme_2024}. These deformations are often slow and shallow, yet could profoundly affect cell behavior. Nevertheless, the feedback between such deformations and active stresses remains poorly understood. Cells do not passively respond to geometry but actively remodel it, and this dynamic interplay may alter their collective state. For instance, proliferating tissues and migrating cell clusters generate forces that deform their surroundings~\cite{matoz-fernandez_wrinkle_2020, winkler_concepts_2020}, raising the question: Can small, activity-driven deformations provide a mechanism for symmetry-breaking in contractile active systems?

\begin{figure*}
\includegraphics[width=1\textwidth]{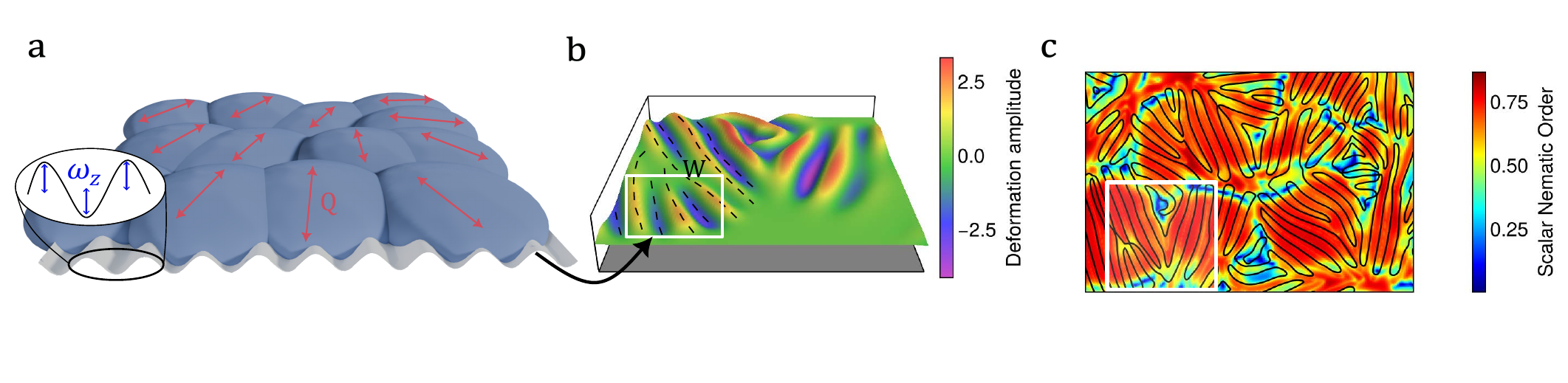}
\caption{\label{fig:snapshot} {\bf Emergence of oreintational order in a contractile system coupled to a deformable substrate.} (a) A schematic representation of the nematic orientation $\mathbf{Q}$, and the substrate out-of-plane deformation orientation $\omega_z$. (b) Zoomed-in 3D representation of the deformation amplitude field of the region highlighted by the white box in (c), with highlighted $\mathbf{W}$ showing relation between the substrate $\omega_z$ and $\mathbf{W}$. (c) Nematic order parameter in a typical simulation, with the black contour marking zero out-of-plane deformation.}
\end{figure*}

In this Letter, we demonstrate that coupling contractile active nematics to deformable environments can induce robust orientational order even in the absence of thermodynamic ordering mechanisms. Using a well-established model of active nematohydrodynamics coupled to a compliant thin film, we uncover a sharp, deformation-mediated transition from isotropic to nematic states. This mechanism stabilizes extended nematic correlations in contractile systems (Fig.~\ref{fig:snapshot}b,c ) and leads to distinct wrinkling morphologies depending on the nature of the active stress. Our findings reveal a generic feedback loop between activity, order, and deformation, providing a new route to spontaneous symmetry breaking in contractile active matter and directly connecting to biological systems where topographical cues guide cellular organization.

A key example of such guidance is the phenomenon of \textit{contact guidance}, where cells orient and migrate in response to substrate features~\cite{weiss_cellular_1959,curtis_topographical_1997,leclech_cellular_2020,buskermolen_cellular_2020,sengupta_principles_2021}. In contact guidance, cells align along the topography, such as ridges and grooves~\cite{driscoll_cellular_2014,nguyen_nano_2016,lacroix_emergence_2024}, irrespective of the specific geometric details. To capture how such topographically oriented states can emerge in active matter, we construct a minimal model in which nematics align purely in response to substrate deformations. In this way, the model provides a coarse-grained realization of contact guidance. Unlike the curvature-alignment mechanism~\cite{bell_active_2022}, which distinguishes between positive and negative curvature at the single-cell scale, our coupling reflects the mesoscale phenomenon of contact guidance, where alignment occurs along the principal axis of deformation. This establishes substrate-mediated alignment as a generic, collective pathway to order in contractile active matter, distinct from curvature-driven effects.

In our model,the topographical landscape, set by the out-plane deformation $\omega_z$, aligns the nematic. We construct a tensor, $\mathbf{W}=\vec{n}_w\otimes \vec{n}_w - \mathbf{I}/2$ via a headless unit vector $\vec{n}_w$, denoting the direction of the topography as shown in Fig.~\ref{fig:snapshot}a,b . For any point in the deformation field, there exists two principal directions or curvature, which are the eiganvectors of the Hessian matrix of the field $\omega_z$. The axis along the deformation, identified as the direction of least curvature is chosen to be $\vec{n_w}$ as this is the direction that cells (and in our model, nematics) tend to align with under contact guidance. In order to induce such alignment in our model, we introduce the term $ k_w (\mathbf{W} : \mathbf{Q})$ into the nematic free energy. 

\begin{equation}
F = \int dV \left[ \frac{B}{4} \mathbf{Q}^4 + \frac{k_n}{2} (\nabla \mathbf{Q})^2 + k_w (\mathbf{W} : \mathbf{Q}) \right],
\end{equation}

The new term penalizes misalignment of the nematic director with the direction of the topography, W. This free energy additionally excludes any standard Landau–de Gennes terms that thermodynamically drive ordering, thereby favouring the isotropic state in the absence of activity ($\zeta=0$) and substrate alignment ($k_w=0$). Through this work we use $k_w \geq 0$ with $\vec{n}_w$ as the eigenvector corresponding to smallest eigenvalue as convention; in principle we would have the same final alignment preference by instead choosing $k_w \leq 0$ with $\vec{n}_w$ as the largest eigenvector, see End Matter for more details.

The nematic dynamics follow standard nematohydrodynamics~\cite{doostmohammadi_active_2018, ramaswamy_mechanics_2010}, and the system is driven out of equilibrium by active stresses $\Pi^{\text{active}} = -\zeta \mathbf{Q}$, where $\zeta$ denotes the activity strength, distinguishing extensile $\zeta > 0$ from contractile $\zeta < 0$ systems. These stresses act on a deformable substrate modelled as a thin elastic sheet on a viscoelastic foundation, described by Föppl–von Kármán theory~\cite{huang_dynamics_2006, landau_theory_2012, ni_modeling_2011}. The total stress $\sigma_{ij} = -p \mathbf{I} + (\mathbf{\nabla v})_{ij} +\Pi^{\text{nematic}}_{ij} + \Pi^{\text{active}}_{ij} + \Pi^{\text{deform}}_{ij}$, with $p$ the pressure and $(\mathbf{\nabla v})_{ij}$ the strain rate tensor, drives both out-of-plane ($\omega_z$) and in-plane deformations ($\omega_x, \omega_y$):
\begin{align}
\frac{\partial \omega_z}{\partial t} &= -K \nabla^4 \omega_z + F \nabla \cdot ( \boldsymbol{\sigma} \cdot \nabla \omega_z) - R \omega_z, \\
\frac{\partial \omega_i}{\partial t} &= \frac{H h_f}{\eta_s} \nabla \cdot \boldsymbol{\sigma} - R \omega_i, \quad i \in {x, y}.
\end{align}
Here, $K$, $F$, and $R$ are fixed elastic and relaxation parameters. The deformations are driven entirely though the the active stress, without which the system equilibrates to a flat, disordered state. The details of the model and explicit definitions of all terms are provided in the End Matter. Minor variations to the model can be found in SI~\cite{SupMat}.

\begin{figure*}
\includegraphics[width=0.9\textwidth]{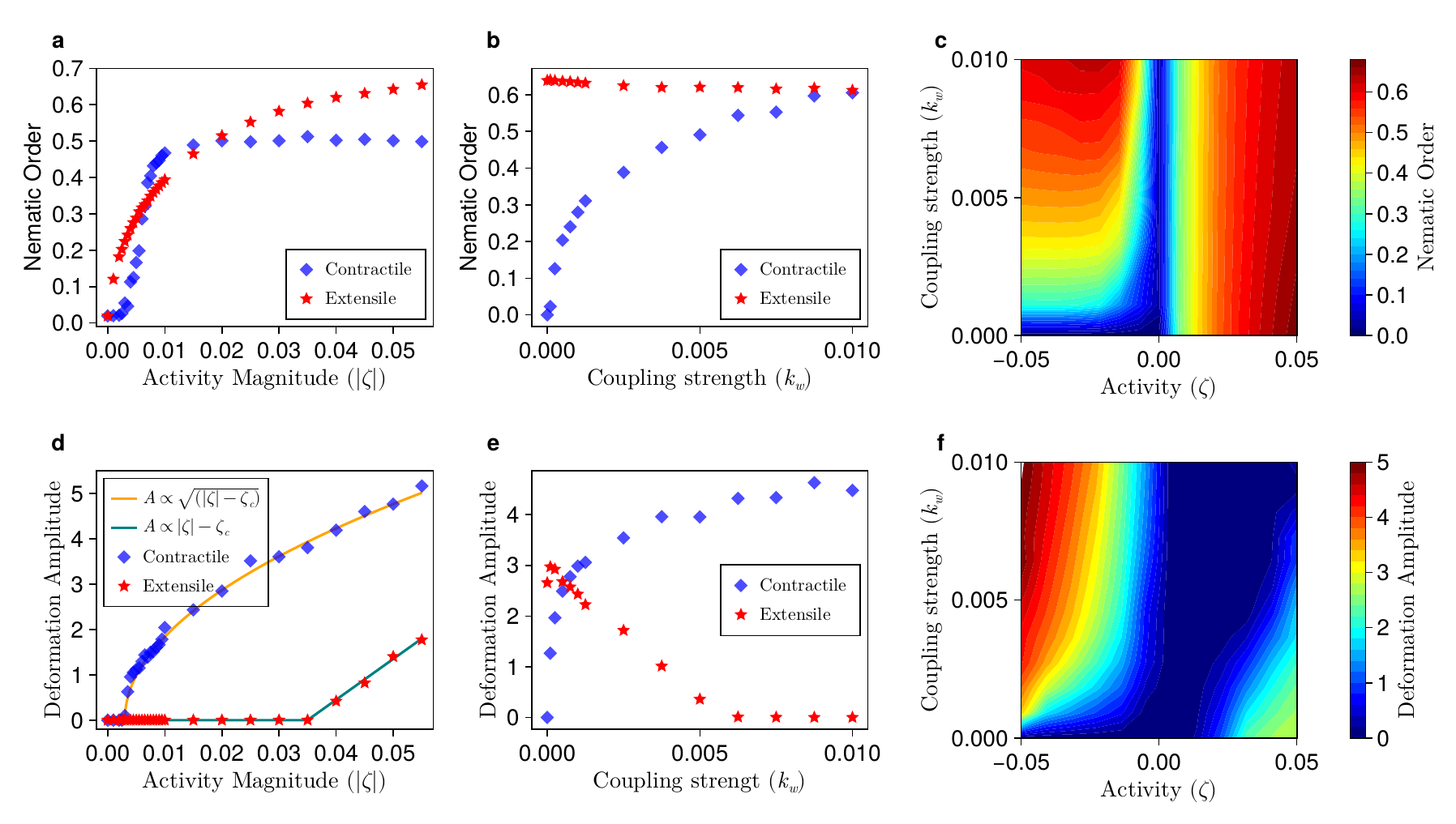}
\caption{\label{fig:onset} {\bf Extensile and contractile activities respond fundamentally differently to deformable substrates.} Nematic order parameter $S$ (a, b) and wrinkle amplitude $A$ (d,e) versus activity $\zeta$ and coupling strength $k_w$. The fixed parameter in (a) and (d) is the coupling strength $k_w = 0.005$; and in (b), (e), the magnitude of the activity $|\zeta| = 0.04$. Phase diagram of $\text{Nematic order}$ (c) and $\text{Amplitude}$ (f).  The growth of $\text{Amplitude}$ with activity $(|\zeta|)$ shows square root (contractile) and linear (extensile) scaling (d), while the dependence of $\text{Amplitude}$ to the coupling strengths $(k_w)$ (e) demonstrates opposite coupling responses for extensile and contractile activities.}
\end{figure*}

To benchmark our model, we fix the flow alignment parameter to $\xi = 1$, a regime where extensile active nematics ($\zeta > 0$) are known to order via flow alignment~\cite{thampi_intrinsic_2015}, while contractile nematics ($\zeta < 0$) remain isotropic. In the presence of coupling to a deformable substrate, for contractile systems, the nematic order parameter $S$ sharply transitions from zero to $S \approx 0.5$ as activity crosses a threshold (Fig.~\ref{fig:onset}a; {\it blue symbols}). This abrupt, plateaued behaviour is in stark contrast with the emergence of order in extensile systems, which occurs at any non-zero activity level (Fig.~\ref{fig:onset}a; {\it red symbols}). Moreover, extensile activity maintains the same ordering regardless of deformation coupling (Fig.~\ref{fig:onset}b; {\it red symbols} $\&$ End Matter Fig.~\ref{fig:kw0}) while in contractile systems the nematic order arises only in the presence of coupling to substrate, and grows gradually with increased coupling strength.(Fig.~\ref{fig:onset}b; {\it blue symbols}).

The ordering for contractile systems coincides with the onset of deformation. Figure~\ref{fig:onset}d shows that the deformation amplitude $A = max(|\omega_z|)$ exhibits a sharp threshold at the activity $\zeta_c$ where order appears. For contractile systems, $A$ scales as $\sqrt{\zeta - \zeta_c}$, consistent with a pitchfork bifurcation, while extensile systems show a linear growth, $A \sim \zeta - \zeta_c$, indicative of a transcritical bifurcation. Notably, the threshold for extensile systems is significantly higher and appears decoupled from ordering, underscoring distinct mechanisms.

Exploring the role of the coupling strength $k_w$ reveals further differences in the mechanisms of order generation in extensile and contractile systems (Fig.~\ref{fig:onset}e). In contractile systems, increasing the coupling $k_w$ amplifies deformation nonlinearly before saturating, reinforcing the geometric feedback. Extensile systems, by contrast, exhibit monotonic suppression of deformation with increasing $k_w$, eventually eliminating substrate modulation altogether (Fig.~\ref{fig:onset}e,f). The full $(\zeta, k_w)$ phase diagram (Fig.\ref{fig:onset}c) reveals a broad regime in which contractile systems become ordered solely due to alignment with small-amplitude substrate deformations. These results establish a purely geometric route to order in contractile active matter - absent in the extensile case - which circumvents traditional thermodynamic ordering that is typically imposed through Landau-de Gennes terms in the free energy.

To understand the mechanisms underlying these behavior, we next examine how ordering and deformation couple differently in the two systems. The decoupling observed in extensile systems, where nematic order persists even when substrate deformations are suppressed, confirms that orientational order arises through flow alignment, as established from prior studies~\cite{saintillan2008instabilities,thampi_intrinsic_2015,santhosh_activity_2020}. Indeed, simulations of extensile system in the absence of hydrodynamics shows that nematic order and deformations are suppressed entirely at all extensile activity levels (End Matter Fig.~\ref{fig:nohydro}). 

\begin{figure}
\includegraphics[width=0.5\textwidth]{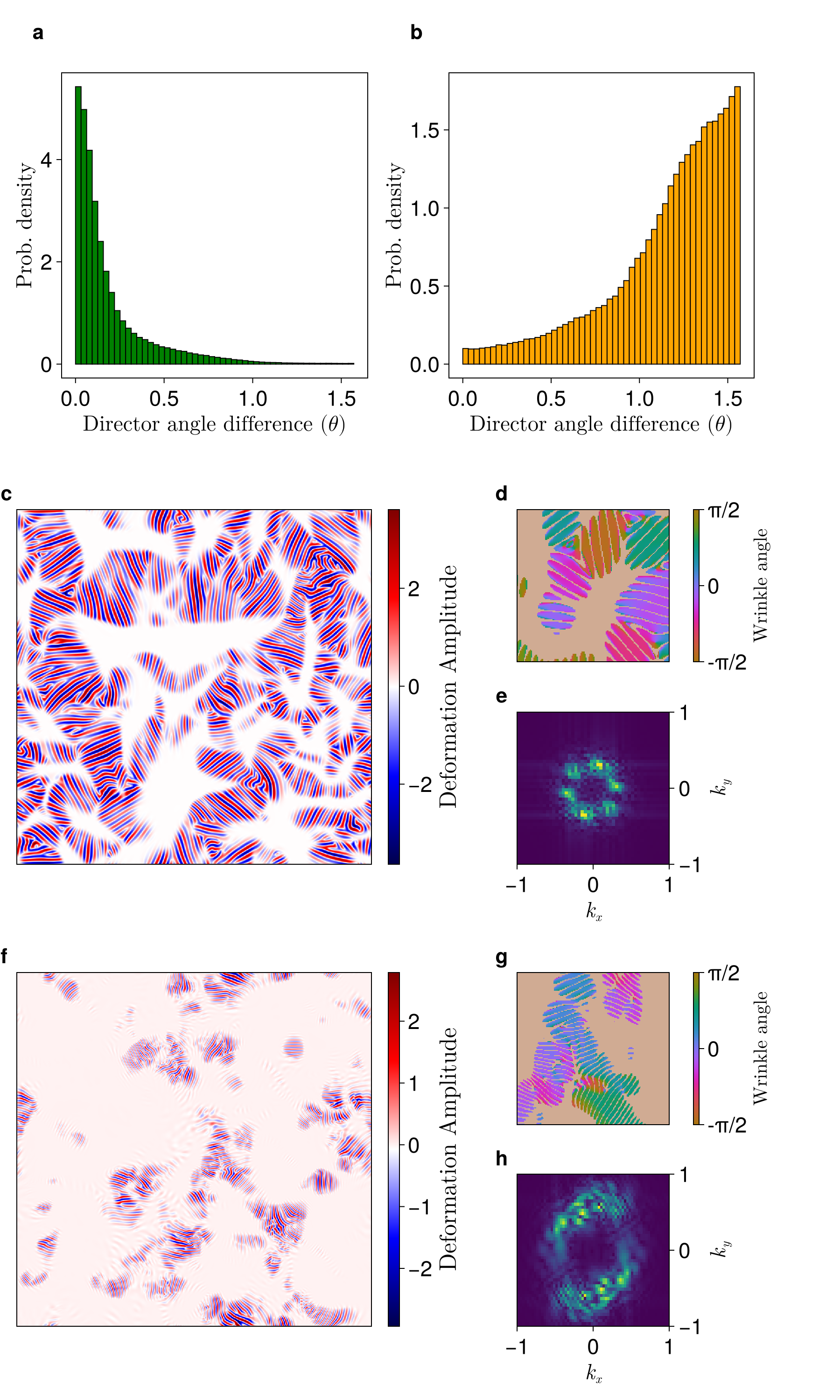}
\caption{\label{fig:local} \textbf{Patterns of deformation in contractile and extensile systems distinguish at the local and mesoscale.} Histograms for contractile (a) and extensile (b) systems show the director preferentially aligns parallel versus perpendicular to the wrinkle. (c) Snapshot of a typical contractile wrinkle pattern. (d) Magnified view of a region in (c), where colour indicates wrinkle orientation, revealing distinct orientation domains, along with (e) the corresponding structure factor. Panels (f)-(h) show equivalent plots for extensile systems, displaying sparse wrinkles and orientation domains with diffuse boundaries. The parameters are fixed at $|\zeta|=0.05$, $k_w=0.005$.}
\end{figure}

In contrast, contractile systems rely on an entirely different mechanism: order emerges only when small-amplitude deformations are present, and increases with the strength of geometric alignment. Indeed, removing hydrodynamics coupling has no effect on the nematic order generation in contractile systems (End Matter Fig.~\ref{fig:nohydro}). This suggests a positive feedback loop between substrate deformation and nematic alignment: contractile activity generates substrate deformations, which in turn align the nematic field, thereby reinforcing the deformation field. We confirm this feedback mechanism by analyzing the local alignment between the nematic director and the deformation orientation. Figure~\ref{fig:local}a,b shows the distribution of the angle $\theta$ between the director and the principal curvature direction. Contractile systems exhibit a strong peak at $\theta \approx 0 \ rad$, indicating parallel alignment, while extensile systems show perpendicular alignment with a peak at $\theta \approx \pi/2 \ rad$.

This perpendicular alignment is additionally responsible for the suppression of deformations in the extensile regime as seen in Fig.~\ref{fig:onset}e. While extensile stresses generate deformations perpendicular to the director the substrate coupling energetically favours parallel alignment of Q with W; thereby creating a negative feedback loop between the nematic director and the deformation growth axis.  As coupling strengthens, the director rapidly reorients to match the local deformation axis, but this realignment redirects the active stresses away from the deformations the require sustained stress to grow.  These distinct behaviours further underscore the fundamentally different coupling responses in the two systems.

In addition to having distinct global properties, the mesoscale structure of deformations distinguish extensile and contractile systems. In contractile regimes, the deformations organize into domains of relatively uniform wavelength and consistent directional orientation. These domains are characterized by regions containing deformations with similar orientational properties. Figure~\ref{fig:local}c presents a snapshot of a typical deformation pattern observed in contractile simulations. The magnified view in Fig.~\ref{fig:local}d, displaying a colormap of wrinkle orientations, reveals the intersection of different domains, highlighting their discrete nature with abrupt changes in orientation. The structure factor (Fig.~\ref{fig:local}e), defined as the 2D Fourier transform of the deformation field, further confirms this block-like organizational pattern. Extensile systems, on the other hand, produce deformation patterns with more gradual variations in the orientation (Fig.~\ref{fig:local}f-h).

\begin{figure}
\includegraphics[width=0.5\textwidth]{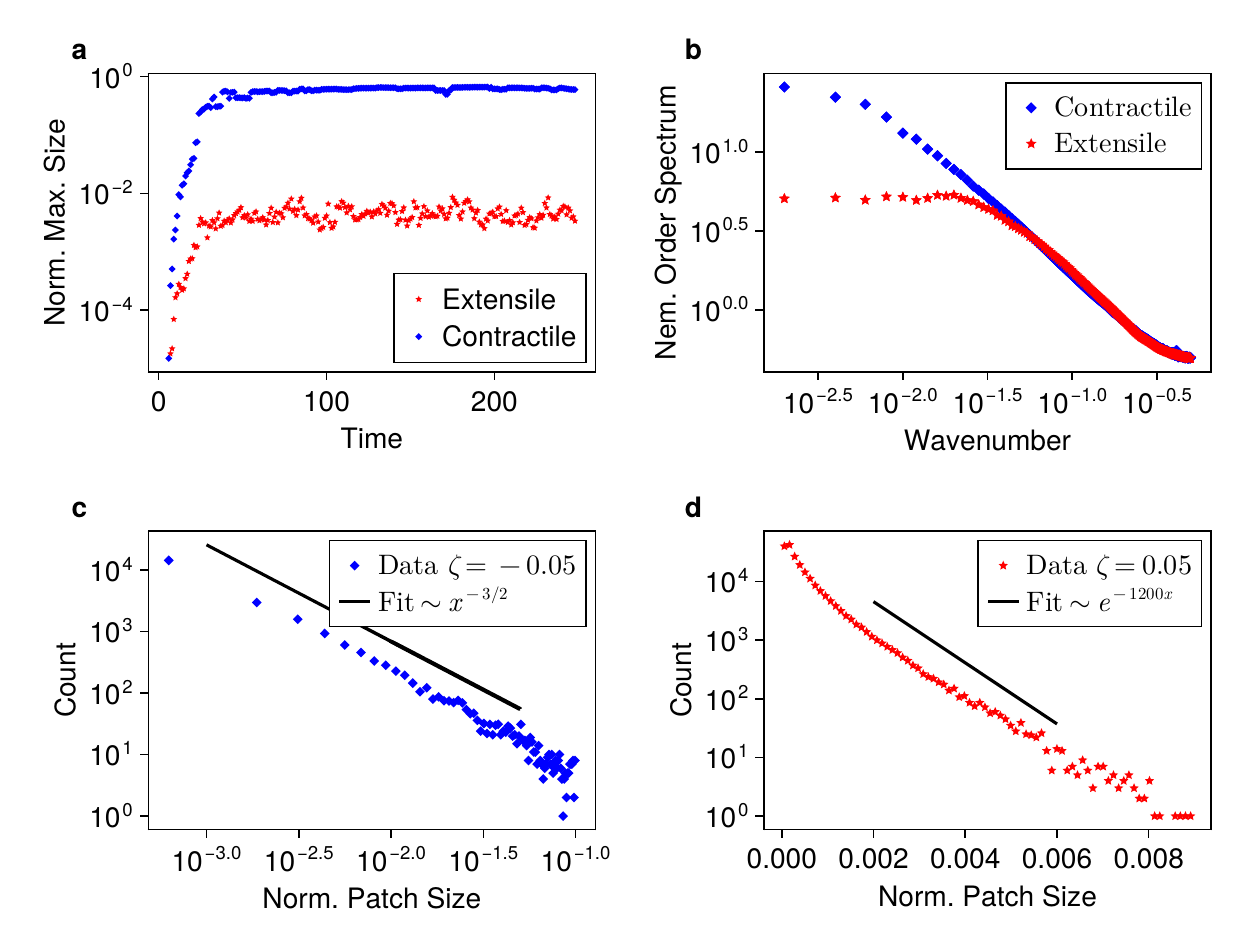}
\caption{\label{fig:consiquence} \textbf{Algebraic correlations and system-spanning power-law deformations emerge in contractile active nematics.} (a) Time series of largest wrinkle domain size normalized by system size, showing contractile systems form system spanning patterns. (b) Power spectrum of the nematic order. (c,d) Patch size distributions, the number of deformation domains of a given area divided by the system size, reveal that patches in contractile systems exhibit power-law scaling (c), large patches in extensile systems follow an exponential distribution (d). The parameters are fixed at $|\zeta|=0.05$, $k_w=0.005$. }
\end{figure}

Furthermore, contractile systems demonstrate the ability to form system-spanning wrinkle domains. Figure~\ref{fig:consiquence}a presents a time series comparison of the largest domain size. The plot shows the area of the largest cluster normalized by the system size, revealing how, in the contractile case, the deformations quickly span almost the entire system within the same time frame that extensile systems take to reach their steady-state size. This is particularly striking, as the contractile patterns are nearly two orders of magnitude larger than the extensile ones. 

To quantify the evolution of deformation domains, we compile histograms of the number of patches versus their area and observe that extensile systems exhibit an exponential decay at large sizes, whereas contractile systems display a scale‐free power law $P(A)\propto A^{-3/2}$ over more than a decade.  In contractile systems, the absence of a characteristic scale and the $3/2$ exponent signal a merger‐dominated growth. Small contractile regions coalesce hierarchically into larger domains, and once a domain exceeds a critical size its concentrated contractile stresses recruit additional material in a positive feedback loop, rendering these patches self‐sustaining, long-lived, and correlated over long times.  Conversely, in extensile materials patches grow only through transient local activity bursts and lack the “pinning” contractile forces needed to stabilize boundaries, so growth beyond a characteristic area becomes exponentially unlikely, with large patches that appear relaxing quickly to inflate the population of small fragments relative to a pure exponential trend. These behaviours are visible in the supplementary videos 1 and 2. 

These structural distinctions suggest a difference in the form of nematic order in the system. The decoupled extensile system is expected to exhibit typical short-range order, as discussed earlier. In contrast, the strong coupling in contractile systems, driven by a positive feedback loop, suggests that the alignment of deformations into large, coherent domains could support extended nematic correlations across the system. To probe large-scale correlations, we measured the power spectrum of the scalar order parameter. Contractile systems exhibit a negative spectral exponent at small wavevectors, consistent with algebraic decay of correlations over the accessible length scales (Fig.~\ref{fig:consiquence}b).

Together, our measurements put forward the following mechanism for order generation on compliant substrates: Extensile nematics achieve order through the balance of active stress against viscous resistance, as established previously in the absence of substrate coupling\cite{thampi_intrinsic_2015,santhosh_activity_2020}. When hydrodynamics are removed, both orientational order and substrate deformations vanish, confirming that active flows, and not geometry, drive pattern formation in this regime. Because the director realigns locally to the surrounding shear, deformation “patches” remain small, transient, and exponentially distributed. By contrast, contractile systems cannot leverage flow alignment and instead harness substrate compliance to break symmetry. A nascent deformation created by contractile stress reorients nearby nematic particles towards its principal curvature, which in turn amplifies the deformation - closing a positive feedback loop that drives patches to coalesce and grow without hydrodynamic mediation. The resulting power-law distribution of patch sizes and the emergence of algebraic decay in the nematic spectrum demonstrate that geometry-driven self-organization provides a sufficient mechanism for robust order in contractile active matter.

The emergence of extended nematic correlations in our model, mediated by feedback between active stresses and substrate deformations, connects to a broader set of ordering phenomena recently reported. In particular, active nematics confined to fluid interfaces have been shown to sustain both polar and nematic order by coupling to the surrounding three-dimensional medium~\cite{maitra_two-dimensional_2023}, while strong substrate friction~\cite{thijssen_role_2020} or intrinsic chirality~\cite{li_pattern_2020} can organize defects or vortices into ordered arrays. These advances demonstrate that coupling to additional physical fields or constraints provides robust pathways to stabilize correlations that would otherwise be short-ranged. In this context, our work highlights substrate compliance as a distinct route: deformation-mediated feedback in contractile systems can generate collective alignment and emergent order, placing it alongside defect ordering and active interface ordering as part of a growing landscape of mechanisms through which active matter achieves large-scale organization.

Future work may explore the change in wrinkle patch distribution, characterize local orientational domains, or investigate long-range nematic correlations in experiments. Experimentally, established in~vitro systems such as fibroblast monolayers on soft substrates, epithelial monolayers, or actomyosin networks on compliant gels provide natural platforms to test the qualitative predictions of our model. Key signatures to probe include deformation-mediated pattern formation, alignment of nematics with substrate anisotropy, and possible enhancement of order on compliant versus stiff environments. Although quantitative parameter matching may require careful interpretation across different experimental contexts, these qualitative behaviours should be directly accessible.

\begin{acknowledgments}
It is a pleasure to acknowledge helpful conversations with Kristian Thijssen, and valuable feedback on the manuscript from Tianxiang Ma. A. D. acknowledges funding from the Novo Nordisk Foundation (grant No. NNF18SA0035142 and NERD grant No. NNF21OC0068687), Villum Fonden (Grant no. 29476), and the European Union (ERC, PhysCoMeT, 101041418). Views and opinions expressed are however those of the authors only and do not necessarily reflect those of the European Union or the European Research Council. Neither the European Union nor the granting authority can be held responsible for them.
\end{acknowledgments}

\section*{Data and materials availability}
The data that support the findings of this article are available at~\cite{data_availability_ref}

\bibliography{references} 

\newpage
\section*{End Matter}

\appendix

\section{Model details}

\subsection{Föppl--von Kármán}

The full Föppl--von Kármán model for a bi-layer of a thin elastic layer over a viscoelastic layer has been used in several studies~\cite{huang_dynamics_2006}. The equations of deformation are derived via a minimisation of bending and stretching energies. The equations are:

\begin{eqnarray}
\frac{\partial \omega_z}{\partial t} &=& -K \nabla^4 \omega_z + F \nabla \cdot (\boldsymbol{\sigma} \cdot \nabla \omega_z) - R \omega_z, \\
\frac{\partial \omega_i}{\partial t} &=& \frac{H h_f}{\eta_s} \nabla \cdot \boldsymbol{\sigma} - R \omega_i, \quad i \in \{x, y\}.
\end{eqnarray}

The stress and strain are defined as:

\begin{eqnarray}
    \sigma_{ij} &=& \Pi^{\text{external}}_{ij} + 2\mu_f\epsilon_{ij} + \frac{\nu_f}{1-\nu_f}\epsilon_{kk}\delta_{ij}, \\
    \epsilon_{ij} &=& \frac{1}{2} \left( \frac{\partial \omega_i}{\partial x_j} + \frac{\partial \omega_j}{\partial x_i} \right) + \frac{1}{2} \left( \frac{\partial \omega_z}{\partial x_i} \frac{\partial \omega_z}{\partial x_j} \right),
\end{eqnarray}

where $\nu_f$ is the Poisson ratio, $\mu_f$ is the shear modulus of the thin film with thickness $h_f$, and $\Pi^{\text{external}}_{ij}$ is the additional in-plane stress exerted by the active fluid. The parameters $K$, $F$, and $R$ are defined as:

\begin{eqnarray}
    K &=& \frac{(1-2\nu_s)\mu_f h_f^3 H}{12(1-\nu_s)(1-\nu_f)\eta_s}, \\
    F &=& \frac{(1-2\nu_s)h_fH}{2(1-\nu_s)\eta_s}, \\
    R &=& \frac{\mu_R}{\eta_s}
\end{eqnarray}

where $\nu_s$, $H$, and $\eta_s$ are the Poisson ratio, thickness, and viscosity of the viscoelastic substrate, respectively.

\subsection{Active Nematics}

Active nematic fluids are complex fluids consisting of rod-like molecules that utilize energy at the microscopic scale. They are described through the interaction of an orientation field and a velocity field. The dynamics of the orientation field are governed by the order parameter tensor \( \mathbf{Q} = S(\mathbf{n} \otimes \mathbf{n} - \mathbf{I}/2) \), where \( \mathbf{n} \) is the director field and \( S \) is the scalar order parameter.
\begin{equation}
    \partial_t \mathbf{Q} + \mathbf{v} \cdot \nabla \mathbf{Q} - \mathbf{S} = \Gamma \mathbf{H},
\end{equation}
where \( \mathbf{H} = -\left[\frac{\delta F}{\delta \mathbf{Q}} - \frac{\mathbf{I}}{2} \mathrm{tr}\left(\frac{\delta F}{\delta \mathbf{Q}}\right)\right] \) is the molecular field driving relaxation toward the free energy minimum of \( \mathbf{Q} \). The free energy \( F \) is given by:
    \begin{equation}
        F = \int \mathrm{d}V \left[ \frac{B}{4} \mathbf{Q}^4 + k_n (\nabla \mathbf{Q})^2 + k_w (\mathbf{W} : \mathbf{Q}) \right].
    \end{equation}

The term \( \mathbf{S} \), often called the co-rotation term, describes the rotation of the director in response to flow via the shear rate \( \mathbf{E} \) and vorticity tensor \( \mathbf{\Omega} \):
\begin{equation}
\begin{split}
    \mathbf{S} = (\xi \mathbf{E} + \mathbf{\Omega})\left(\mathbf{Q} + \frac{\mathbf{I}}{2}\right) 
    + \left(\mathbf{Q} + \frac{\mathbf{I}}{2}\right)(\xi \mathbf{E} - \mathbf{\Omega})
     \\ 
     - 2\xi \left(\mathbf{Q} + \frac{\mathbf{I}}{2}\right) \mathbf{Q} : \nabla \mathbf{v}.
     \end{split}
\end{equation}

The velocity field follows the incompressible Navier-Stokes equations:
\begin{align}
    \nabla \cdot \mathbf{v} &= 0, \\
    \rho \left( \partial_t \mathbf{v} + \mathbf{v} \cdot \nabla \mathbf{v} \right) &= \nabla \cdot \mathbf{\Pi},
\end{align}
where \( \mathbf{\Pi} \) is the stress tensor:
\begin{equation}
    \mathbf{\Pi} = -p\mathbf{I} + 2\eta \mathbf{E} - \zeta \mathbf{Q} + \mathbf{\Pi}^{\text{nematic}}.
\end{equation}

The nematic stress \( \mathbf{\Pi}^{\text{nematic}} \) is given by:
\begin{equation}
    \mathbf{\Pi}^{\text{nematic}} = 
    \begin{aligned}[t]
    & 2\xi \left(\mathbf{Q} + \frac{\mathbf{I}}{2}\right)(\mathbf{Q} : \mathbf{H}) \\
    &- \xi \mathbf{H} \cdot \left(\mathbf{Q} + \frac{\mathbf{I}}{2}\right) 
    - \xi \left(\mathbf{Q} + \frac{\mathbf{I}}{2}\right) \cdot \mathbf{H} \\
    &- \nabla \mathbf{Q} \frac{\delta \mathcal{F}}{\delta \nabla \mathbf{Q}} 
    + \mathbf{Q} \cdot \mathbf{H} - \mathbf{H} \cdot \mathbf{Q}.
    \end{aligned}
\end{equation}

The nematohydrodynamic equations are formulated in Cartesian coordinates rather than on the curved surface geometry. This approximation is justified because the characteristic length scale of substrate deformations is significantly smaller than both the nematic coherence length and the hydrodynamic screening length. Under this scale separation, the substrate curvature varies smoothly on the scales relevant to nematic and hydrodynamic processes, allowing the local Cartesian approximation to remain valid.

\subsection{Parameter space and simulation specifics}

Simulations were done using the hybrid lattice Boltzmann method~\cite{lbm}, a grid of $L_x=Ly = 1048$, with the following parameters fixed (in lattice Boltzmann units): $K_n$ = 0.005, $\xi$ = 1.0,  $\Gamma$ =  1.0, $ \rho = 40$, $ B $  = 0.01, $\eta$ =  0.5, $\eta_s$ = 0.5, $H = 10$, $h = 0.5$, $K$ = 0.01, $F = 0.35$, $R = 0.0005$.


\subsection{Alignment coupling}

The topographic alignment term $k_w (\mathbf{W}:\mathbf{Q})$ penalizes deviations between the nematic director and the substrate's principal direction $\vec{n}_w$, defined as the eigenvector of the smallest eigenvalue of the Hessian of $\omega_z$ (i.e., the direction of least curvature). The Hessian matrix $\mathbf{Hess}$ of the deformation field is given by:

\begin{equation}
\mathbf{Hess} = \begin{pmatrix}
\partial^2_x \omega_z & \partial_x\partial_y \omega_z \\
\partial_x\partial_y \omega_z & \partial^2_y \omega_z
\end{pmatrix},
\end{equation}

where $\partial^2_i \omega_z$ denotes second spatial derivatives. The sign convention $k_w \geq 0$ ensures alignment with $\vec{n}_w$; an equivalent description could use $k_w \leq 0$ with $\vec{n}_w$ as the \textit{largest} eigenvector (maximal curvature direction).  $k_w = 0$ recovers the traditional active nematics. 

Additional simulations reveal qualitatively different behaviors for static versus dynamic substrates. For contractile systems ($\zeta<0$) with strong $k_w$ coupling to a \textit{static} curved surface, the director field locks uniformly to the curvature direction, forming defect free, equilibrium like configurations without flows. This contrasts sharply with the dynamic, self-sustained patterns observed when the substrate evolves (as described in the main text). Extensile systems ($\zeta>0$), however, remain largely unaffected by static W. 

These results, as well as results for minor variations in the Free energy are shown in the SI. We find that substrate induced contractile ordering persists when the coupling term has a threshold of $\omega_z = 1e^{-3}$, the LdG potential also includes a positive quadratic term ($F = 0.5C(Tr(Q^2) + 0.5Tr(Q^2)^2)$), and when the coupling term depends linearly on the amplitude.

\section{Hydrodynamics is not required for contractile nematic order}

\begin{figure}
    \centering
    \includegraphics[width=0.9\linewidth]{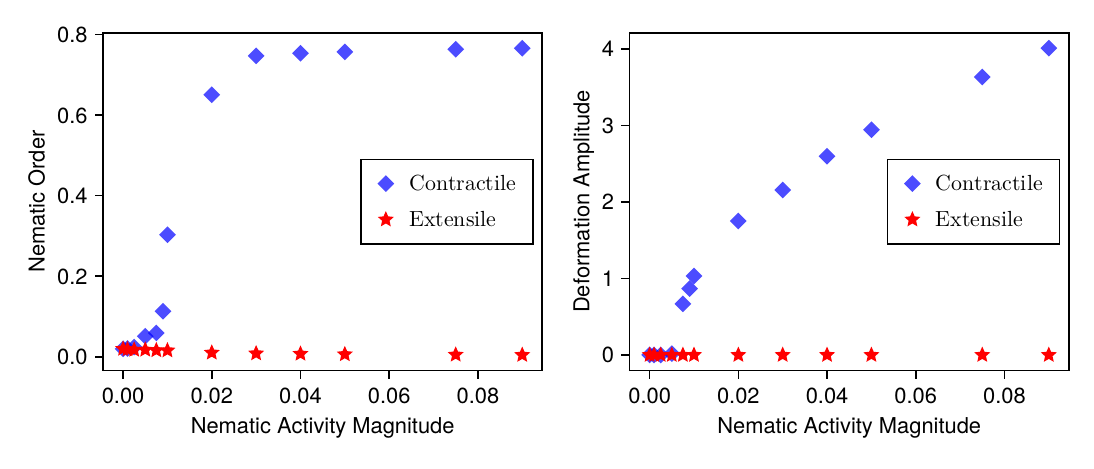}
    \caption{Contractile activity generates oreintational order and deformations without hydrodynamics, while extensile activity shows no ordering in the absence of hydrodynamics, with $k_w=0.005$}.
    \label{fig:nohydro}
\end{figure}

We find that in the absence of hydrodynamics, only contractile activity is seen to generate both nematic order and out of plane deformations. Similar to Fig.~\ref{fig:onset}, the nematic order shown by contractile activity rises abruptly before reaching a plateau (Fig.~\ref{fig:nohydro}) Extensile activity, in contrast, does not show any order at all. 

\section{Order generation in extensile systems does not depend on substrate coupling} 

\begin{figure}
    \centering
    \includegraphics[width=0.9\linewidth]{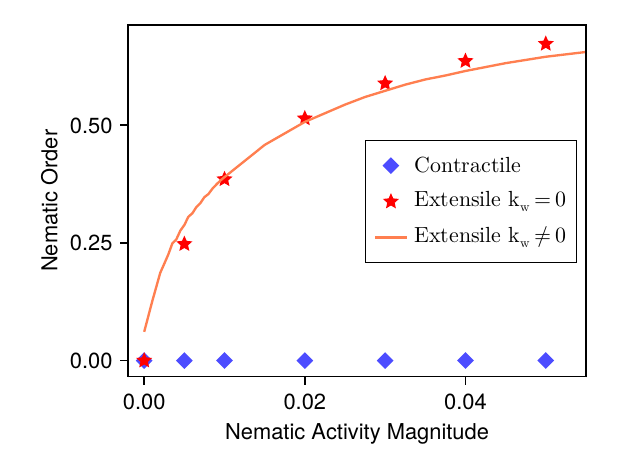}
    \caption{The strength of coupling, $k_w = 0.005$ vs $k_w = 0$, to the substrate does not change the nematic ordering for extensile activity, but contractile system shows no order with $k_w=0$.} 
    \label{fig:kw0}
\end{figure}

Here, we demonstrate that the nematic ordering for extensile activity is independent of the substrate coupling strength (Fig.~\ref{fig:kw0}, red symbol and line). This observation has two important implications. By being independent, we not only show that the ordering mechanism is purely flow driven, but also demonstrate that the active stress increases faster than linearly as a function of $\zeta$ for extensile activity, as $\Pi^{act}\sim\zeta Q$; $Q\sim S$, where the order $(S)$ is non-linear.  

The contractile case however shows no ordering at all and remains at $S=0$ for all activities (Fig.~\ref{fig:kw0}, blue symbols).

\end{document}